\documentclass[twocolumn,showpacs,amsmath,amssymb,prl,superscriptaddress]{revtex4}
\usepackage{amsmath}
\usepackage{graphicx}
\usepackage{bm}
\usepackage{color,ulem} 

\newcommand{\e}{{\text{e}}}

\newcommand{\p}{\text{ph}}
\newcommand{\ee}{\text{e-e}}
\newcommand{\qeq}{\text{neq}}
\newcommand{\eq}{\text{eq}}
\newcommand{\ep}{\text{e-ph}}
\newcommand{\T}{\text{T}}

\begin{document}

\title{Nonequilibrium electrons in tunnel structures under high-voltage injection}

\author{N.B. Kopnin}
\affiliation{Low Temperature Laboratory, Helsinki University of
Technology, P.O. Box 5100, 02015 TKK, Finland}
\affiliation{L. D. Landau Institute for Theoretical Physics, 117940 Moscow, Russia}
\affiliation{Argonne National Laboratory, Argonne, IL 60439, USA}

\author{Y.M. Galperin}
\affiliation{Department of Physics, University of Oslo, PO Box
1048 Blindern, 0316 Oslo, Norway}
\affiliation{ A.F. Ioffe
Physico-Technical Institute of Russian Academy of Sciences, 194021
St. Petersburg, Russia}

\author{J. Bergli}
\affiliation{Department of Physics, University of Oslo, PO Box
1048 Blindern, 0316 Oslo, Norway}

\author{V.M. Vinokur}
\affiliation{Argonne National Laboratory, Argonne, IL 60439, USA}

\date{\today}

\begin{abstract}
We investigate electronic distributions in nonequilibrium tunnel
junctions subject to a high voltage bias $V$ under competing
electron-electron and electron-phonon relaxation processes. We
derive conditions for reaching quasi-equilibrium  and show that, though the
distribution can still be thermal for low energies where the
rate of the electron-electron relaxation exceeds significantly the
electron-phonon relaxation rate, it develops a power-law tail at
energies of order of $eV$. In a general case of comparable
electron-electron and electron-phonon relaxation rates, this tail
leads to emission of high-energy phonons which carry away most of
the energy pumped in by the injected current.
\end{abstract}

\maketitle

Two important classes of out-of-equilibrium phenomena in
mesoscopic systems can be identified: (i) those that are described
by \textit{quasi-equilibrium} quasiparticle distributions where
electrons and phonons have well defined, although in general
different, temperatures and (ii) those where quasiparticle
distributions deviate significantly from their equilibrium form
and where the notion of temperature cannot be introduced. The
correct interpretation of the data observed in the particular
experiment requires the proper attribution of the specific
phenomenon to either of these classes. Superconducting mesoscopic
structures such as tunnel and Josephson junctions are among those
that are being investigated most intensively. The common tools for
inferring information about nonequilibrium states in such systems
are studies of the electron-phonon interaction
rates~\cite{kaplan76,reizer89,day03,kozorezov01,barends08,RothwarfTaylor67}
and of the electron-phonon energy
relaxation~\cite{gantmakher74,roukes85,wellstood94,Girvin96,giazotto06,karvonen07,Timofeev09}.
A standard setup is an island connected to leads (they all can be
superconducting or normal) via tunnel contacts and driven out of
equilibrium by strong electron injection under bias voltages $V$
such that $eV$ significantly exceeds both the temperature and the
superconducting gap $\Delta$. In many experiments (see, e.g.,
Ref.~\onlinecite{Timofeev09}), a quasi-equilibrium
distribution is created with an electronic temperature $T_\e$
essentially higher than the bath temperature $T_\p$ kept low by
efficient cooling. These temperatures determine the energy
transfer from electrons to the phonon bath. The quasi-equilibrium
establishes when a big difference between fast electron-electron,
$\gamma_{\ee}$, and slow electron-phonon, $\gamma_{\ep}$,
relaxation rates exists at energies $\epsilon \sim T_\e$. These
conditions are usually fulfilled in aluminum samples at sub-Kelvin
temperatures.

The more common is the situation, however, where, in contrast to
Al samples, the ratio $\gamma_\ee/\gamma_\ep$ at such temperatures
is not very high (as illustrated, e.g., by the data in Ref.
\cite{kaplan76}). In this Letter we investigate formation of the
electronic distribution in  the general case and show that for
moderate ratios $\gamma_\ee/\gamma_\ep$ the conditions of
quasi-equilibrium can be easily violated. The distribution formed
under a high-voltage injection can be characterized by an effective electronic
temperature $T_{\e}> T_{\p}$ only within a low-energy region
around $\epsilon \sim T_\e$ provided $\gamma_{\ee}(T_\e)$ is much
larger than $\gamma_{\ep}(T_\e)$. We find that at energies of
order of $eV\gg T_\e$, the distribution has a long power-law tail,
which crucially changes the transport properties of the entire
electronic system having much lower effective
temperature $T_\e$. Since the tail energies exceed
$\Delta$, our results are general and apply to both normal
and superconducting junctions. While the derived behavior somewhat
resembles the well known electron runaway in
semiconductors~\cite{Frolich}, it is drastically different from
such nonequilibrium effects as superconductivity stimulation
\cite{IvlevEliashberg}, nonequilibrium proximity effect in
Josephson junction (see, e.\,g.,~\cite{Kozorezov08} and references
therein), etc., where the deviation from equilibrium is maximal
right in the energy range near $\Delta$.

We consider the energy exchange between
the electron and phonon subsystems relevant to a rich
variety of
experiments~\cite{gantmakher74,roukes85,wellstood94,Girvin96,giazotto06,karvonen07,Timofeev09}.
Since $\gamma_{\ep}(\epsilon)$ is well known to grow faster than
$\gamma_{\ee}(\epsilon)$ with increasing energy, there exists
certain energy,  $\epsilon^*$, at which these two rates match. One
would expect that nonequilibrium effects in the energy transfer
are small as long as $T_\e \ll \epsilon^*$. We show that it is
indeed the case if $eV\ll \epsilon^*$ and the electron-electron
interaction dominates in the entire nonequilibrium region.
However, at $eV \gg \epsilon^*$ a crossover from the electron-electron to the electron-phonon
mechanism of relaxation takes place as a function of energy. In this
case the power law tail in the electron distribution leads to
emission of high-energy phonons, which carry away most of the
energy provided by the injected current. Accordingly, the energy
emitted via thermal phonons becomes a much smaller fraction of the
total inserted energy, implying that interpretation of the
experimental data based on the quasi-equilibrium distribution
function with some effective temperature does not apply.


\paragraph{Setup.}

Having in mind mostly superconducting devices, we consider a
junction consisting of a superconducting (or normal)
macroscopic-size island $(i)$ connected via small-area
high-resistance tunnel contacts with two superconducting (or
normal) leads $L_1$ and $L_2$. The bias voltage $V$ is assumed
high in the scale of characteristic energies of the particular
experiment. For the system mentioned above, the proper inequality
is $eV\gg T_\e ,\Delta$. For the energies  $T_\e, \Delta \ll
\epsilon \lesssim eV$, which we are interested in, the
normal-state equations for the electron-phonon, electron-electron
interactions, and the density of states can be used both for a
normal and  superconducting junction. The electron temperature
is assumed uniform along the sample. We use the clean-limit
approximations for $\gamma_\ee (\epsilon)$ and
$\gamma_\ep(\epsilon)$ because impurity scattering does not
significantly renormalize the electron-electron or electron-phonon
interactions for energies of interest. The influence of the
impurity scattering on the electron-phonon relaxation is
controlled by the parameter $q \ell$ where $\ell$ is the
electronic mean free path and $q=\epsilon /\hbar s$ is the wave
vector of an emitted phonon with energy
$\epsilon$~\cite{Altshuler78,ReizerSergeev85,sergeev00}. Taking
sound velocity $s\sim 5000$ m/s and $\ell \sim 20$ nm as in Al
samples of~\cite{Timofeev09} we get $q\ell\sim (0.5~{\rm
K}^{-1})~\epsilon /k_B$. Therefore, at $\epsilon/k_B \sim eV/k_B
\gg 1$ K the clean limit is appropriate. The impurity-induced
renormalization of the electron-electron interaction
\cite{AltshulerAronov} becomes important for energies $\epsilon
\lesssim \hbar^3/\tau(p_F\ell)^2 \sim \epsilon_F(\hbar/p_F\ell)^3$
and can also be neglected for $\epsilon \sim eV$.
Under these assumptions we can derive a linear equation for the
distribution function and solve it exactly for the most relevant
situations.

We consider a symmetric structure with voltages at the leads $
V_{L_1}=-V_{L_2}=V/2 $ such that the chemical potential of the
island $\mu_i=0$ by symmetry, while for the leads
$\mu_{L_1}=-\mu_{L_2}\equiv \mu_L =-eV/2$. It is convenient to
write kinetic equations for odd and even components of the
distribution function $n(\epsilon)$ defined as $f_1(\epsilon) =
n(-\epsilon) -n(\epsilon)$, and $f_2(\epsilon) =1- n(\epsilon)
-n(-\epsilon)$. If the leads are in thermal equilibrium at
temperature $T_L$, we have $f_1^{(L_1)}= f_1^{(L_2)}\equiv
f_{1}^{(L)}$ and $f_2^{(L_1)}=-f_2^{(L_2)} \equiv f_2^{(L)}$ where
\[
f_{1,2}^{(L)} =\frac{1}{2}\left[ \tanh \frac{\epsilon
-eV/2}{2T_{L}}\pm \tanh \frac{\epsilon +eV/2}{2T_{L}}\right]\ .
\]
In what follows we assume that the leads and the phonon bath are
at zero temperature, $T_L=T_\p =0$, and that all the emitted phonons are
immediately removed from the sample due to ideal heat contact to
the substrate.

One can check that the even component of the distribution in the
island vanishes by symmetry, $f_2 (\epsilon)=0$ . The kinetic
equation for the odd component is
\begin{equation}
J_1^\T+J_1^{(\e)}+J_1^{(\p)}=0 \label{kin-eq-total}
\end{equation}
where $J_1^\T$, $J^{(\e)}_1$, and $J^{(\p)}_1$ are, respectively,
the tunnel, electron-electron, and electron-phonon collision
integrals in the island. The latter describe relaxation of the
distribution driven out of equilibrium by the tunnel
source~\cite{VoutilHeikkKopnin05}
\begin{equation}
J_1^\T= -4\eta \,[f_1 -f_1^{(L)}]  \label{tun-norm}
\end{equation}
that contains the distribution in the leads, $f_1^{(L)}$, and in
the island, $f_1$.  The (identical) tunneling
contacts are characterized by an effective tunneling rate $ \eta
=(4\nu e^2 \Omega R)^{-1} $ where $\nu\equiv \nu(E_F)$ is
the normal density of states in the island, $\Omega$ is its
volume, and $R$ is the contact resistance. In what follows we
consider high contact resistances, i.\,e., small $\eta$ (the
estimate will be given later).

\paragraph{Electron-phonon relaxation.}

For small $\eta$, the distribution function determined by
Eq.~(\ref{kin-eq-total}) is close to the thermal,
$f_1(\epsilon)\approx \tanh (\epsilon/2T_{\e})$, with a certain
electronic temperature $T_{\e}$. For energies $\epsilon \sim T_\e$
the deviation from quasi-equilibrium is negligible. However, at
$\epsilon \gg T_{\e}$ where $\tanh (\epsilon/2T_{\e})\approx {\rm
sign}(\epsilon)$, the injection-induced deviation becomes
essential. We put $f_1(\epsilon)=\tanh
\left(\epsilon/2T_{\e}\right)+\delta f(\epsilon)$ where $|\delta
f|\ll 1$. At $T_\p=0$, the electron-phonon collision
integral~\cite{kopnin} in the island vanishes for
$f_1(\epsilon)={\rm sign}(\epsilon)$. Therefore, neglecting
exponentially small terms we get for energies $|\epsilon|,
|\epsilon +\omega| \gg T_{\e}$ and $\epsilon
>0$ (see also Ref. \cite{Volkov})
\begin{equation}
J_1^{(\p)}= -
\frac{\gamma_{\ep}(T_\e)}{T_\e^3}\left[\frac{\epsilon^3\delta
f(\epsilon)}{3}   - \int_0 ^{\infty}\! \! \! \! d\omega \,
\omega^2 \delta f(\epsilon+\omega)\right] \label{eq1}
\end{equation}
where $ \gamma_{\ep}(T_\e)= \pi \lambda_{\ep} T_\e ^3/2\hbar
(sp_F)^2 $ is the electron-phonon relaxation rate at the
electronic temperature $T_\e$, while $\lambda_{\ep}$ is the
interaction constant. We use $T_\e$ simply as a convenient energy
scale. In fact, the ratio $\gamma_\ep(T_\e)/T_\e^3$ is independent
of $T_\e$.

\paragraph{Electron-electron interaction.}

The electron-elec\-tron collision integral satisfies
the energy conservation law,
$\int_{-\infty}^{\infty}  \epsilon J_1^{(\e)}(\epsilon)\, d\epsilon
=0$~\cite{kopnin}.
For large energies $|\epsilon|,
|\epsilon_1|,|\epsilon_2|,|\epsilon_3| \gg T_{\e}$ and $\epsilon
>0$ it has the form
\begin{equation}
J_1^{(\e)}=-\frac{\gamma_{\ee}(T_\e)}{T_\e^2}\left[
\frac{\epsilon^2 \delta f(\epsilon)}{2} - 3\! \int_0^\infty \! \!
\! d\omega \, \omega \,  \delta f(\epsilon+\omega)\right]\, .
\label{J-ee}
\end{equation}
Here $ \gamma_\ee(T_\e)=\pi \lambda_\ee T_\e^2/8\hbar E_F $ is the
electron-electron relaxation rate at $T_\e$, $\lambda_\ee$ is the
interaction constant. The ratio $\gamma_\ee(T_\e)/T_\e^2$ is
independent of $T_\e$.


\paragraph{Distribution function.}

The electron-electron relaxation dominates at very low energies,
where the distribution has a thermal form with an electronic
temperature $T_{\e}$. At higher energies, a deviation from thermal
behavior develops  due to the reduced role of the
electron-electron interaction. Consider $\epsilon>0$. Since
$f_1(\epsilon)\approx 1$ for $\epsilon \gg T_{\e}$ the tunnel
collision integral Eq.~(\ref{tun-norm}) becomes $ J_1^\T= -4\eta
\, \Theta \left(eV/2 -\epsilon \right)$. This form of the
injection term suggests that at $\epsilon \gg T_{\e}$,
  $ \delta f(\epsilon)=
-\phi(\epsilon)\,  \Theta \left(eV/2-\epsilon \right)$. Using Eqs.
(\ref{tun-norm}) -- (\ref{J-ee}), the kinetic equation
(\ref{kin-eq-total}) becomes
\begin{eqnarray}
&&\frac{1}{\epsilon_\p^3}\left[\frac{\epsilon^3}{3}\phi(\epsilon)
- \int_0 ^{eV/2-\epsilon} \! \! \!d\omega \, \omega^2
\phi(\epsilon+\omega)\right] \nonumber \\&& \quad
+\frac{1}{\epsilon_\e^2}\left[ \frac{\epsilon^2}{2} \phi(\epsilon)
- 3\int_0 ^{eV/2-\epsilon}\! \! \!d\omega \, \omega \,
\phi(\epsilon+\omega)\right]=1. \label{eq-3}
\end{eqnarray}
Here $ \epsilon^{3}_{\p}=4\eta T_\e^3/\gamma_{\ep}(T_\e)$ and
$\epsilon^{2}_{\e}=4\eta T_\e^2/\gamma_{\ee}(T_\e) $. Equation
(\ref{eq-3}) has a characteristic energy scale
\begin{equation}
\epsilon^* =\epsilon_{\p}^3/\epsilon_{\e}^2\sim
T_\e[\gamma_{\ee}(T_\e)/\gamma_{\ep}(T_\e)] \ , \label{charact-en}
\end{equation}
such that $\gamma_\ep(\epsilon^*)=\gamma_\ee(\epsilon^*)$.
Therefore, the electron-phonon interaction dominates for $\epsilon
\gg \epsilon^*$ while the electron-electron interaction dominates
for $\epsilon \ll \epsilon^*$. As it is explained earlier, we
consider here the situation when $\gamma_{\ee}(T_\e)\gg
\gamma_{\ep}(T_\e)$, thus $\epsilon^* \gg T_\e$.

The integral equation (\ref{eq-3}) can be transformed into a
differential one by triple differentiation over energy, which can
then be easily analyzed and solved numerically. The boundary
conditions are obtained by putting $\epsilon =eV/2$ at each step.
In the situations of dominant electron-electron relaxation,
$\gamma _{\ep}\to 0$ (i.\,e., $\epsilon_\p \rightarrow \infty$), it
can be reduced to the second-order differential equation for a
function $\phi_\e$ determined entirely by the electron-electron
interaction
\begin{equation}
\frac{d^2}{d\epsilon^2}\left[ \epsilon^2 \phi_\e(\epsilon)\right]
=6 \phi_\e(\epsilon)\, , \label{eq-gen3}
\end{equation}
with the boundary conditions $\epsilon^2\phi_\e(\epsilon)
=2\epsilon_\e^2$ and $(d/d\epsilon)[\epsilon^2\phi_\e(\epsilon)]
=0$ at $\epsilon =eV/2$. Solution of Eq.~(\ref{eq-gen3}) with
these boundary conditions is
\begin{equation}
\phi_\e = \frac{4\epsilon_\e^2 }{5}\left[
\frac{3}{2}\frac{(eV/2)^2}{\epsilon^4}+\frac{\epsilon}{(eV/2)^3}\right]\, .
\label{phi-ee}
\end{equation}
Fast electron-phonon relaxation, $\gamma _{\ep} \to \infty$, leads
to
\begin{equation}
\frac{d^3}{d\epsilon^3}\left[\epsilon ^3 \phi_\p
(\epsilon)\right]+6\phi_\p(\epsilon)=0
\end{equation}
with the conditions  $\epsilon ^3 \phi_\p (\epsilon)
=3\epsilon_\p^3$, $(d/d\epsilon)[\epsilon ^3 \phi_\p (\epsilon)]
=(d^2/d\epsilon^2)[\epsilon ^3 \phi_\p (\epsilon)] =0$ at
$\epsilon =eV/2$. The solution is
\begin{eqnarray}
\phi_{\p}(\epsilon)&=&\frac{18}{11}\left( \epsilon_{\p}/\epsilon
\right)^3\mathcal{F}\left(eV/2\epsilon \right)\, , \label{phi-eph} \\
\mathcal{F}(x) &\equiv& x+\frac{1}{x^2}\left[ \frac{5}{6}\cos
(\sqrt{2}\ln x) + \frac{\sqrt{2}}{3}\sin (\sqrt{2}\ln x)\right] \,
. \nonumber
\end{eqnarray}
It was obtained in \cite{Volkov} and used in~\cite{Kozorezov1} for
studies of nonequilibrium created by absorption of a high-energy
photon.

At comparatively low voltages, $ eV \ll \epsilon^* $, the
electron-electron interaction dominates in the entire energy range
$0<\epsilon <eV/2$ and the distribution function obeys
Eq.~(\ref{phi-ee}). At high voltages, $eV \gg \epsilon^*$, one can
discriminate between two regions with different relaxation
mechanisms with a crossover between them at $\epsilon \sim
\epsilon^*$. For $0<\epsilon\ll \epsilon^*$ the electron-electron
interaction dominates. For $\epsilon^* \ll \epsilon <eV/2$ the
electron-phonon interaction wins, and the distribution function
approaches Eq.~(\ref{phi-eph}).
In both cases the electronic distribution has a long power-law
nonequilibrium tail $ \phi(\epsilon ) \sim
(\tilde{\epsilon}/\epsilon)^4$ at energies $T_{\e} \ll \epsilon
\ll eV$. For $ eV \ll \epsilon^* $, one has $\tilde \epsilon =
(\epsilon_\e eV)^{1/2}$, while $\tilde \epsilon = (\epsilon_\p^3
eV)^{1/4}$ for $ eV \gg \epsilon^* $. The deviation from
equilibrium becomes of the order unity for $\epsilon \lesssim
\tilde \epsilon$. Thus the low energy distribution can be thermal
only if $\tilde{\epsilon} \ll T_\e$. This condition reads
\begin{equation}
eV/\epsilon^* \ll \xi\cdot \max\{\xi, 1\}, \quad \xi \equiv
T_\e^2/\epsilon_\e\epsilon^*\, . \label{quasieq}
\end{equation}
Therefore, one can interpret the data inferred from the experiment
in terms of a quasi-equilibrium electronic temperature $T_\e$ only
if the condition (\ref{quasieq}) is fulfilled. Otherwise, the
distribution is not thermal even at small $\epsilon$; instead, the
scale of its variation, and thus the apparent ``temperature'', is
determined by $\tilde \epsilon$.

\paragraph{Energy balance.}

Even if the quasi-equilibrium condition (\ref{quasieq}) holds, the
long power-law tail in the distribution can strongly influence
electronic processes at a much lower temperature $T_\e$. An
important example is the electron-phonon energy relaxation. Using
Eq.~(\ref{kin-eq-total}) one finds the energy balance per unit
volume of the island
\begin{equation}
2\nu \int _{0}^{eV/2}\epsilon J_1^\T\, d\epsilon +2\nu \int
_{0}^{eV/2}\epsilon J_1^{(\p)}\, d\epsilon =0 \, . \label{bal1}
\end{equation}
The electron-electron collision integral vanishes due to the
energy conservation.
With Eq.~(\ref{eq1}) this gives
\begin{equation} \label{balance-gen}
P_\eq (T_\e)+ P_\qeq = P_V \, .
\end{equation}
Equation (\ref{balance-gen}) determines $T_{\e}$ as a function of
the injected power $ P_V = V^2/4R\Omega$. The latter is obtained
from the first integral in Eq.\ (\ref{bal1}) under the assumption
that $eV \gg T_\e$. This is the power (per unit volume) deposited
into the island in a setup with two contacts of total resistance
$2R$ under the total voltage bias $V$. It is half of the total
input energy, the other half of which goes into the leads. The
l.h.s. of Eq.~(\ref{balance-gen}) comes from the second integral
in Eq.\ (\ref{bal1}) and consists of the energy transfers, $P_\eq$
and $P_\qeq$, to zero-temperature phonon bath by thermalized and
by nonequilibrium electrons, respectively. The energy transferred
by thermal electrons, $P_\eq(T_{\e})$, is determined by the
electron-phonon interaction integrated over the low-energy domain
$\epsilon \gtrsim T_\e$ with $f_1=\tanh(\epsilon /2T_\e)$
neglecting the nonequilibrium correction. For a clean normal
metal, $P_\eq(T_{\e})=\Sigma T_\e^5$ where
$ \Sigma\approx 78 \nu \lambda/(sp_F)^2\hbar $~\cite{wellstood94}. For clean
superconductors, $P_\eq(T_{\e})$ was calculated
in~\cite{Timofeev09}.

The energy transferred by non-thermal electrons is
\begin{equation} \label{Pneq}
P_\qeq=\frac{8\nu \eta}{\epsilon_\p^3}\!\int _{0}^{eV/2} \! \! \! \!
\! \!
\epsilon d\epsilon \!
\left[\frac{\epsilon^3\phi(\epsilon)}{3}
- \! \int_0^{eV/2-\epsilon} \! \! \! \! \!\! \! d\omega \, \omega^2
\phi(\epsilon+\omega)\right].
\end{equation}
The main contribution here comes from $\epsilon \sim eV$.

For voltages $eV\ll \epsilon^*$ when the electron-electron
interaction always dominates, the non-thermal contribution can be
calculated using Eq.~(\ref{phi-ee}). The $\epsilon^{-4}$ term in
Eq.\ (\ref{Pneq}) is cancelled out leading to $
P_\qeq=(eV/3\epsilon^*)P_V $, and Eq.\ (\ref{balance-gen}) yields
\begin{equation}
P_\eq (T_{\e})=P_V\left[1-  eV/3\epsilon^*\right]\ .
\label{power1}
\end{equation}
The second term in the brackets is a small correction: Almost all
the injected energy is absorbed and then transmitted to the
phonon bath by thermal electrons.

The situation is totally different at higher bias voltages, $eV
\gg \epsilon^*$. In this case almost all the injected power is
absorbed by high energy electrons, and $T_\e$ is less sensitive to
$V$. Using $\phi_{\p}(\epsilon)$ from Eq.~(\ref{phi-eph}), which
satisfies the kinetic equation without the electron-electron
collision integral, the energy balance Eq.~(\ref{balance-gen})
becomes
\begin{eqnarray*}
&&P_\eq(T_{\e})=\frac{8\nu \eta}{\epsilon_\p^3}\int _{0}^{eV/2} \! \! \!
\epsilon\, d\epsilon
\left(\frac{\epsilon^3}{3}[\phi_\p(\epsilon)-\phi(\epsilon)]
\right. \nonumber \\
&&\quad - \left. \int_0
^{eV/2-\epsilon}\! \! \! \omega^2 d\omega \,
[\phi_\p(\epsilon+\omega)-\phi(\epsilon+\omega)]\right).
\end{eqnarray*}
The function $\phi(\epsilon)$ satisfies the full kinetic
equation~(\ref{eq-3}). The main contribution to the integral comes
from $\epsilon\sim eV \gg \epsilon^*$ where Eq.~(\ref{eq-3})
coincides with that for small electron-electron interaction within
the accuracy $\epsilon^*/eV$. Therefore $ (\phi_{\p}-\phi)/\phi_\p
\sim \epsilon^*/eV$. This results in a slower dependence of
$P_\eq$ on the bias voltage,
\begin{equation}
P_\eq(T_{\e})\sim P_V(\epsilon^*/eV) =
\nu \eta \epsilon^* eV \, .\label{power2}
\end{equation}
Expressions (\ref{power1}) and (\ref{power2}) are the main result
of our Letter. They reveal a crossover from $\propto V^2$ to
$\propto V$ behavior that occurs at $eV
\approx \epsilon^*$. This is confirmed by 
numerical solution of the kinetic equation~(\ref{eq-3}).

\paragraph{Discussion.} The data for aluminum
samples~\cite{kaplan76,giazotto06,ee-eph-rate} suggest $ \gamma_\e
(T_c)\sim 10^8~{\rm s}^{-1}$, $ \gamma_\p(T_c) \sim 10^6~ {\rm
s}^{-1}$, while $ \eta = 10~{\rm s}^{-1}$ for the samples of
Ref.~\cite{Timofeev09}. Therefore, $ \epsilon^*/T_c \sim 10^2 $.
For voltages $eV\sim 10^2\,T_c$ used in \cite{Timofeev09}, the
electron-electron interaction always dominates, while the
deviation from equilibrium remains small. Thus the
quasi-equilibrium conditions are well satisfied.

However, the above situation is rather an exception to the rule --
it is in a sense unique to aluminum samples. In other materials
the electron-phonon relaxation rates are normally much
larger~\cite{kaplan76}. Consequently, their ratios to the
corresponding electron-electron relaxation rates are not as low as
in Al. Therefore, the crossover from electron-electron to
electron-phonon dominated relaxation occurs at much lower
energies, and should be observable in experiments. Although the
deviation from the thermal distribution could be still small
within the low-energy range $\epsilon \sim T_\e$, the high-energy
tail of the distribution at $\epsilon \sim eV$ becomes essential
for the energy exchange between the electron and phonon
subsystems. The injected power will then be mostly absorbed by
high-energy phonons with energy $T_\e \ll \epsilon \lesssim eV$.

To summarize, we have established the quasi-equilibrium condition
Eq.\ (\ref{quasieq}), and demonstrated that the energy exchange
rate $P_{\eq}(T_\e)$ between thermal electrons and the phonon bath
experiences a crossover from Eq.\ (\ref{power1}) to Eq.\
(\ref{power2}) as a function of the bias voltage. The crossover
energy determines the relative strength of the electron-electron
and electron-phonon interactions.

\acknowledgments We thank J. Pekola for many stimulating
discussions. This work was supported by the Academy of Finland, by
the Russian Foundation for Basic Research under grant
09-02-00573-a, by Deutsche Forschungsgemeinschaft within GK 638,
by the U.S. Department of Energy Office of Science under Contract
No. DE-AC02-06CH11357, and by Norwegian Research Council through
STORFORSK program. YG is thankful to the Ben Gurion University of
Negev and to the Weizmann Insitute of Science for hospitality.

\end{document}